\documentclass[]{spie}  

\usepackage{amsmath,amsfonts,amssymb}
\usepackage{graphicx}
\usepackage{hyperref}
\hypersetup{colorlinks,allcolors=blue}

\title{TCSpy: Multi-telescope Array Control Software for 7-Dimensional Telescope (7DT)}

\author[a]{Hyeonho Choi}
\author[a]{Myungshin Im}
\author[b]{Ji Hoon Kim}
\affil[a]{SNU Astronomy Research Center, Astronomy Program, Seoul National University, 1 Gwanak-ro, Gwanak-gu, Seoul, ,Republic of Korea}

\authorinfo{Further author information: \\Hyeonho Choi: E-mail: hhchoi1022@gmail.com}

\begin{document} 
\maketitle

\begin{abstract}
We introduce a novel software called TCSpy which is designed to efficiently control a multi-telescope array through network-based protocols. The primary objectives of TCSpy include centralized control of the array, support for diverse observation modes, and swift responses to the follow-up observations of astronomical transients. To achieve these objectives, TCSpy utilizes the ASCOM Alpaca protocol in conjunction with Alpyca, establishing robust communication among multiple telescope units. For the practical application of TCSpy, we implement TCSpy within the 7-Dimensional Telescope (7DT). 7DT is a telescope array consisting of 20, 0.5-m telescopes, equipped with 40 different medium-band filters. The main scientific goals of 7DT include detecting the optical counterparts of gravitational-wave sources, identifying kilonovae, and the spectral mapping of the southern sky. Through the integration of TCSpy, 7DT can achieve these scientific objectives with its unique observation modes and rapid follow-up capabilities.
\end{abstract}


\section{Introduction}
\label{sec:intro}  
With the advent of the gravitational-wave (GW) detectors (LIGO\cite{Ligo2015}, Virgo\cite{Virgo2015}, KAGRA\cite{Kagra2021}), a significant number of GW events are expected to be discovered in the near future. While the discovery of these transients alone is scientifically meaningful, additional follow-up observations provide valuable information about the nature of such events\cite{Kilonova_2012}. However, even though a significant number
of telescopes are rapidly responding to interesting transient events for follow-up observations, the number of
observatories is not sufficient. This is particularly true for events with large localization areas and those requiring
spectroscopic observation. For such cases, doing a spectral mapping over the whole localization area can help speed up the identification of the transient of interest.

An efficient solution for both wide-field observation and spectroscopic analysis of the entire image is to utilize a multi-telescope array. With each telescope unit operating independently within the array, the sky localization area can be covered with the combined field of view of all telescopes, without being constrained by a specific tiling pattern. Moreover, this setup enables spectroscopic observation of the entire image by observing the same position with different medium-band filters. To perform this kind of spectroscopic mapping of the sky, our group has developed the 7-Dimensional Telescope (7DT), which is a multiple telescope array compsed of 20 telescopes equipped with 40 different medium filters.

To support various observation modes of 7DT and possibly other multiple telescope systems, we developed a novel software called TCSpy for the operation of a multi-telescope array composed of Astronomy Common Object Model (ASCOM) and PlaneWave instruments (PWI). Utilizing a centralized system, TCSpy supports synchronized operation of multiple telescopes via network-based communication. Currently, TCSpy is specifically designed for the operation of 7DT. This paper demonstrates the software architecture of the TCSpy and its practical application on 7DT. Section~\ref{sec:7-dimensional_telescope} will briefly present the hardware specifications, science goals, and operational requirements of 7DT. Section~\ref{sec:software_architecture} demonstrates network-based hardware control, software design of the TCSpy, and target database. Section~\ref{sec:implementation_of_tcspy_on_7DT} will discuss the implementation of the TCSpy on 7DT for diverse observation modes and robotic observation.

\begin{figure} [ht]
\begin{center}
\begin{tabular}{c} 
\includegraphics[height=8.2cm]{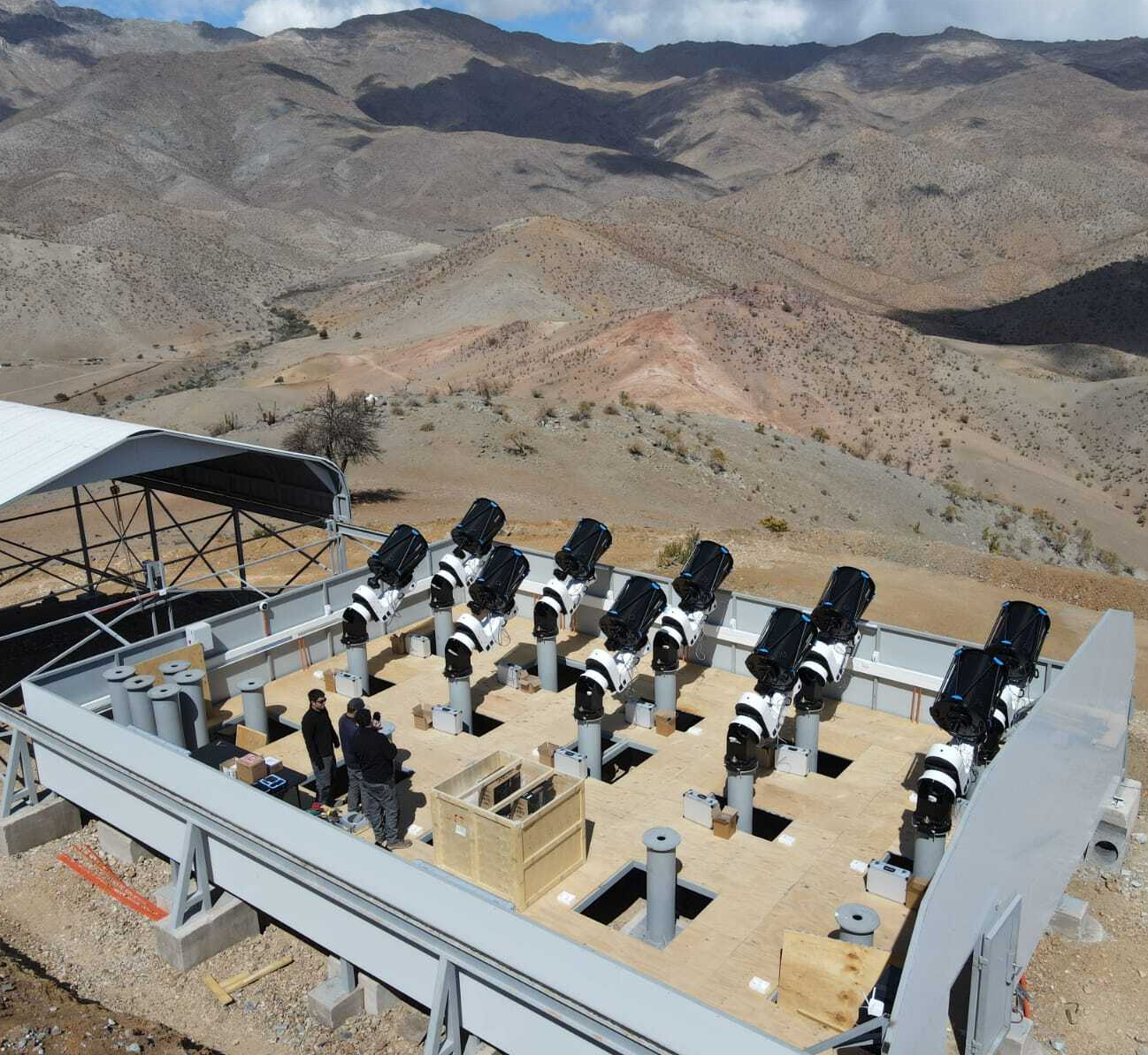}
\includegraphics[height=8.2cm]{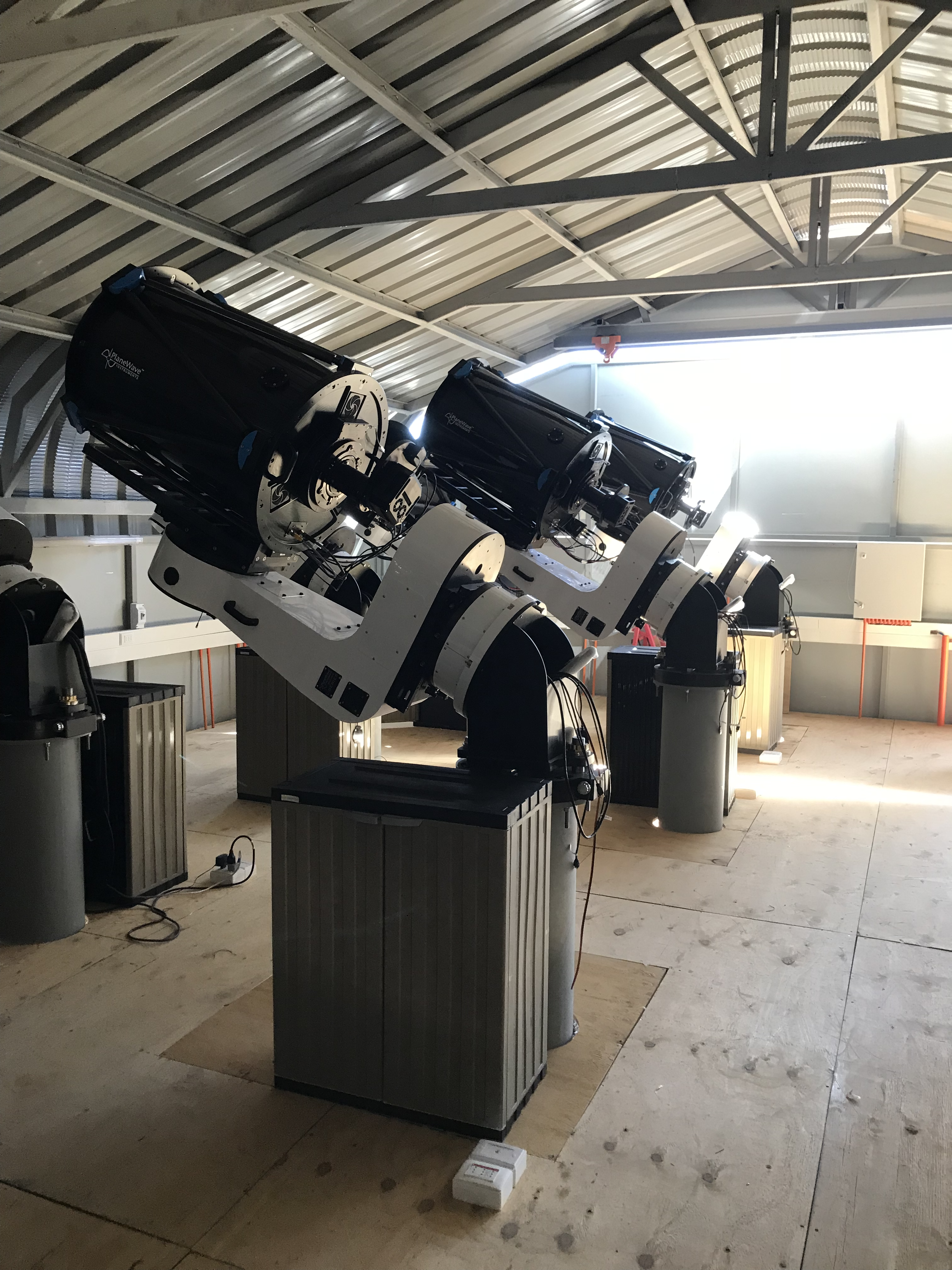}
\end{tabular}
\end{center}
\caption[example] 
{
 (Left) 7DT overview. Currently, 7DT is under construction with 10 of 20 units operational. (Right) Specific view of telescope unit. Each telescope unit is connected to the TCC located below the telescope unit. For more detail of 7DT, see (\url{https://gwuniverse.snu.ac.kr/research/7dt})}
\end{figure} 

\section{7-Dimensional Telescope}
\label{sec:7-dimensional_telescope}
\subsection{Hardware}
\label{subsec:7DT_hardware}
7DT is a multi-telescope array located at the El Sauce Observatory, Chile. It comprises 20 0.5-meter telescopes equipped with 40 different medium-band filters. Each telescope unit includes a fast slewing mount, an F/3 optical tube assembly (OTA) with a 508mm diameter, a focuser, a 61-megapixel CMOS camera, and a 9-slot filter wheel. Each filter wheel is equipped with the Sloan g-, r-, i-band filters, along with four to six 25 nm width medium-band filters. With its large field of view (1.25 square degrees per telescope unit), rapid slewing speed (50 degrees per second), and swift readout speed (less than 2 seconds), 7DT is ideally suited for rapid follow-up observations of transient events. Each telescope unit is operated by an assigned telescope control computer (TCC) with a physical connection. Main Control Computer (MCC) governs all TCC for centralized operation of multiple telescope units. For more detail on hardware control, see Section~\ref{subsec:hardware_control}. Currently, 12 of the 20 units are operational, allowing for telescope standardization and commissioning observations.

\begin{figure} [ht]
\begin{center}
\begin{tabular}{c} 
\includegraphics[height=5.5cm]{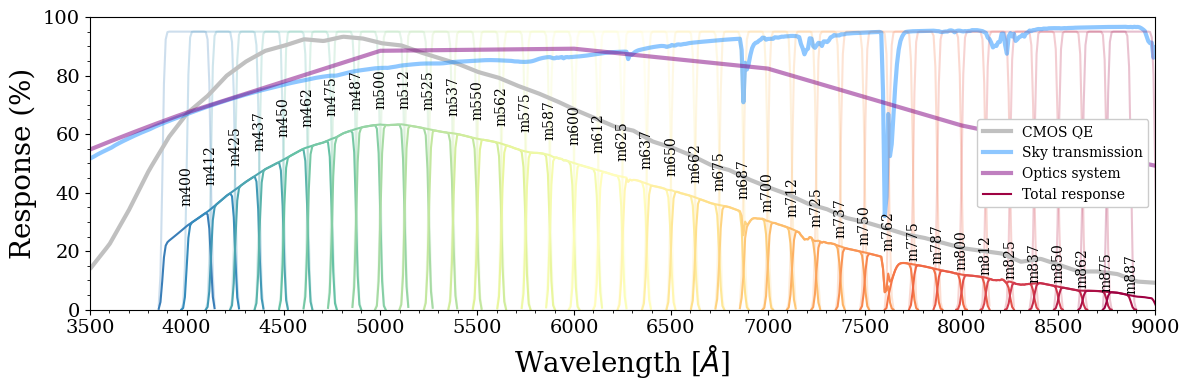}

\end{tabular}
\end{center}
\caption[example] 
{
\label{fig:response_curve} 
The response curve of 40 medium-band filters equipped in 7DT. Each filter is represented by a 3-digit number indicating its central wavelength in nm. By performing observation simultaneously with these medium-band filters, 7DT has the capability of low-resolution spectroscopy from 400nm to 887nm. 
}
\end{figure}

\subsection{Scientific goal}
\label{subsec:7DT_scientific_goal}
One of the main scientific objectives of 7DT is transient optical follow-up observation, primarily used to search for the optical counterpart of GW sources detected by LIGO-Virgo-KAGRA observing run 4\cite{LVK_O4_run}(LVK O4). As the major observational facility in the Center for the Gravitational-Wave Universe and Gravitational-wave Electromagnetic Counterpart Korean Observatory\cite{GECKO2023_Im, GECKO2024_Paek} (GECKO) project, 7DT will trigger target of opportunity (ToO) observations on the sky localization area of GW events and identify optical counterpart in real-time. Moreover, follow-up observation with diverse observation modes on interesting transient events such as supernovae, and high energy sources will be triggered. The details of the observation modes will be discussed in Section~\ref{subsubsec:observation_modes}.

Furthermore, in situations when transient events are not occurring, 7DT will conduct the 7-Dimensional Sky Survey, the spectral mapping survey of the southern hemisphere sky. Depending on a scientific purpose, survey area, and observation cadence, there are three types of surveys: Reference Image Survey (RIS), Wide Field Survey (WFS), and Intensive Monitoring Survey (IMS). The data collected from these surveys will be in the format of data cubes with low-resolution spectroscopy and will be utilized for transient subtraction, photometric redshift calculation, and other analyses.

\subsection{The requirement for the operation}
\label{subsec:The requirement for the operation}
To achieve the scientific goals discussed in Section~\ref{subsec:7DT_scientific_goal}, the basic requirements for 7DT operation are as follows: 1. Support for diverse observation modes. 2. Synchronized operation of multiple telescope units, and 3. Swift responsiveness for ToO observation request.

\subsubsection{Observation modes}
\label{subsubsec:observation_modes}
7DT operates in a variety of observation modes for various astronomical research needs. Taking advantage of multiple telescope units, we need to implement three observing modes: (1) Spectroscopic observation mode (Spec mode); (2) Deep observation mode (Deep mode); and (3) Search observation mode (Search mode).

In Spec mode, each telescope observes the same designated target with different medium-band filters. The default setting of the spec mode should include all medium-band filters from m400 to m887, yet customizable spec modes should be defined as well. The response curves of all medium-band filters are shown in Figure ~\ref{fig:response_curve}.

Deep mode involves the observation of all the telescope units on the same designated target with the same filter. By combining all images obtained from the 20 telescope units, this mode yields a significant enhancement in light-gathering power, equivalent to that of a 2.3-m telescope.

Search mode is essential to rapidly cover extensive sky localization areas such as GW events. In the Search mode, each telescope unit individually observes a different area of the sky. The critical aspect of the Search mode is to avoid assigning duplicate targets to multiple telescope units. 

\subsubsection{Synchronized operation of multiple telescope array}
Rapid spectral changes and weather fluctuation can impact data quality, especially in spec and deep observation modes. Hence, in such observation modes, all telescope units must synchronize their observations on the shared target within a few seconds. Furthermore, in the spec mode, real-time sharing of target observing status and swift observation scheduling is necessary to facilitate swift observations. 

\subsubsection{Swift responsiveness for follow-up observations}
Given that the primary scientific objective of 7DT is to detect transients and conduct follow-up observations, swift ToO observations are essential. Therefore, our primary aim is to ensure that 7DT is autonomously ready for ToO observation within one minute. The rapid follow-up observations will provide crucial information necessary to uncover the nature of transient events.

\section{Software architecture}
\label{sec:software_architecture}
Telescope Control System with Python (TCSpy) is a Python software designed specifically for the synchronized control of ASCOM\cite{ASCOM}-based multi-telescope array. Using ASCOM Alpaca\cite{ASCOM_alpaca} and PWI4 HTTP API\cite{PWI4}, robust communication between telescopes is established with HTTP protocol. Through communication, centralized operation of multiple telescope units becomes feasible, enabling the realization of diverse observation modes with synchronization. Moreover, with the dynamically interactive target database and scheduling system, the telescope array integrated with TCSpy can be fully automatic with the fast response of ToO observation.

\subsection{Hardware control}
\label{subsec:hardware_control}
\begin{figure} [hb]
\begin{center}
\begin{tabular}{c} 
\includegraphics[height=5.5cm]{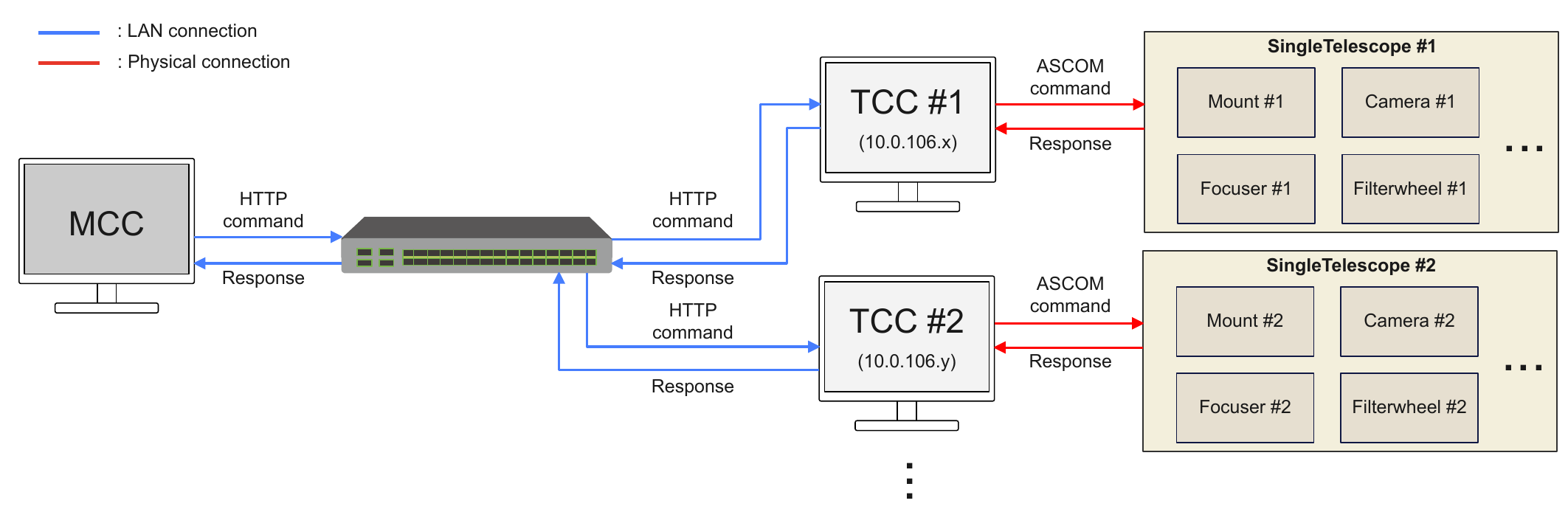}
\end{tabular}
\end{center}
\caption[example] 
{ \label{fig:hardware_control} 
The schema of hardware control in TCSpy. Once a command is triggerd in MCC, the command is conveyed via network to each TCC. Then, each TCC control connected telescope components for operation. 
}
\end{figure}

In TCSpy, all telescope hardware communicates via the network using the HTTP protocol. Each telescope is hosted by its Telescope Control Computer (TCC) through a physical connection. These individually hosted telescopes can be controlled via HTTP commands using their respective IP addresses, enabling centralized control by the Main Control Computer (MCC). The schema of the hardware control of TCSpy via the HTTP protocol is illustrated in Figure ~\ref{fig:hardware_control}. Two different APIs are utilized for controlling telescope components with the HTTP protocol: (1) ASCOM Alpaca, and (2) PWI4 HTTP API.

\subsubsection{ASCOM Alpaca}
ASCOM stands for "Astronomy Common Object Model." It is one of the standards for communication between telescope control software and hardware devices like mounts, focusers, filter wheels, and more. ASCOM provides a standardized interface that allows different software applications to control various devices without needing specific drivers. ASCOM Alpaca is an extension of the ASCOM platform, enabling the control of ASCOM devices over a network. Each ASCOM device is configured as an Alpaca device via an ASCOM Remote Server.  These configured Alpaca devices can be controlled via the host IP address by utilizing the Alpaca device API. TCSpy is built based on a Python library called Alpyca, which allows the control of Alpaca devices from Python. 

\subsubsection{PWI4 HTTP API}
PlaneWave Interface 4 (PWI4) is a device control software provided by PlaneWave. With devices manufactured by PlaneWave, PWI4 offers various functions such as telescope slew, autofocus, and more. Similar to ASCOM Alpaca, PWI4 supports device manipulation via the HTTP protocol using the host IP address. In addition to ASCOM Alpaca API, TCSpy also supports PWI4 HTTP API for mounts and focusers.

\subsection{Software design}
TCSpy is organized into three fundamental types of modules: device, action, and utility modules. These modules play essential roles in device integration, operational actions, and various utility functions. By combining these fundamental modules, users can create applications that enable observers to manage the operation of the multiple telescope array. With the flexibility of Python, each application can also be scheduled for automatic operation. 

\begin{figure} [hb]
\begin{center}
\begin{tabular}{c} 
\includegraphics[height=8cm]{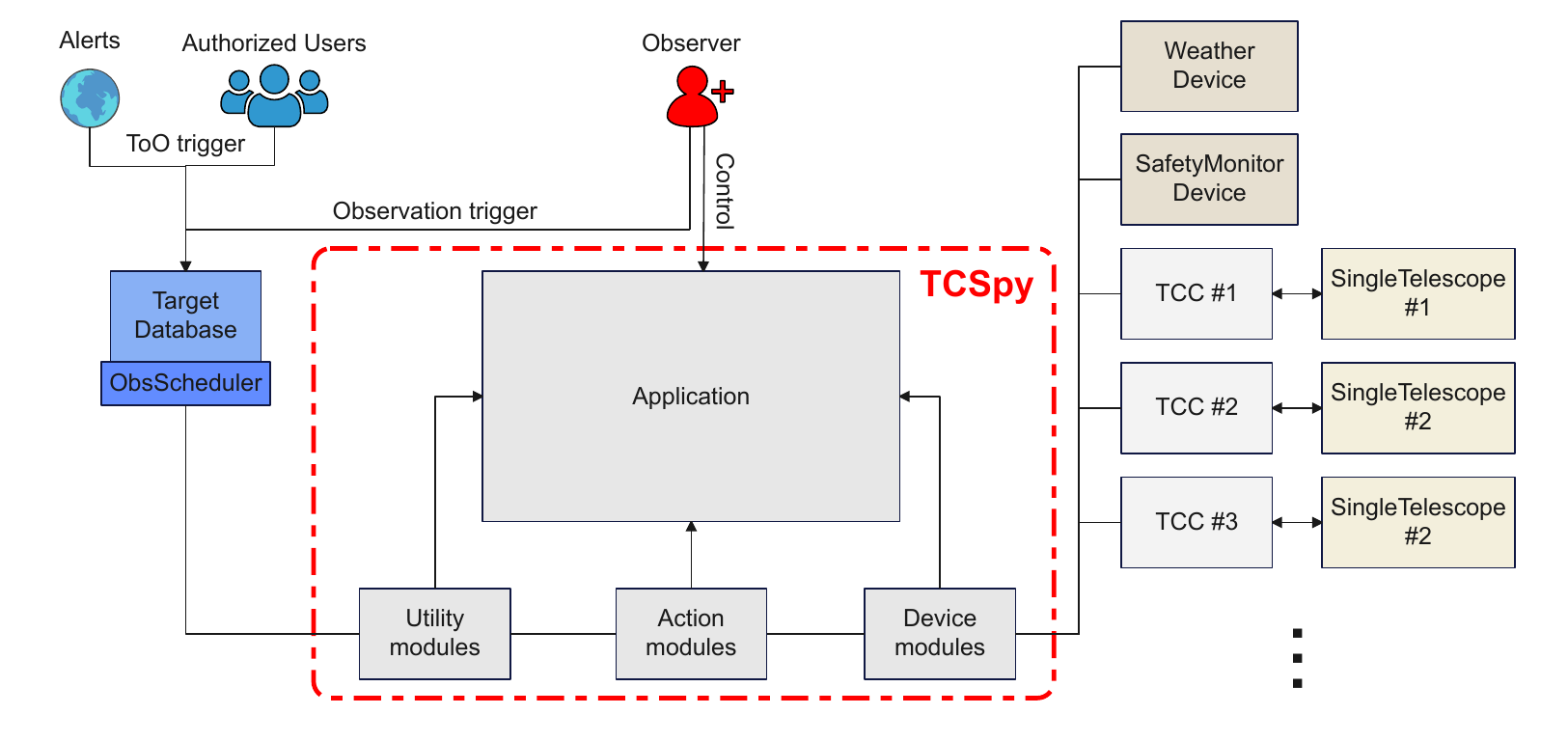}
\end{tabular}
\end{center}
\caption[example] 
{ \label{fig:Software_design}
The software design of TCSpy. There are three basic modules in TCSpy: Device, Action, and Utility modules. With the robust connection with multiple telescope array, all telescopes can be controlled by the centralized control system. By deploying these modules, users can create a desired application with the target database for the robotic observation. 

}
\end{figure}

In the device module, TCSpy establishes connections with the multiple telescope array at various levels. At the lowest level, it connects ASCOM Alpaca devices (such as cameras, filter wheels, etc.) to create the SingleTelescope instance. Each SingleTelescope instance represents a single telescope unit. Multiple SingleTelescope instances are then combined to form MultiTelescope, allowing for the simultaneous observation of multiple telescopes with synchronization across various observation modes.

The Action modules enable the practical operation of both SingleTelescope and MultiTelescope instances. These modules are structured into four levels, depending on the complexity of the required action. For SingleTelescope operations, Level 1 and 2 action modules are employed, while MultiTelescope operations utilize level 3 action modules. All action modules simply provide two interfaces "run" and "abort" for running and aborting an operation. Furthermore, action modules in the levels 1 and 2 can be wrapped by the MultiAction module at the level 0, enabling simultaneous operation across multiple telescopes. All observation modes presented in Section~\ref{subsubsec:observation_modes} are implemented in the action modules. 

The Utility modules provide essential support to TCSpy. These utility modules offer a list of functionalities, including image header control, image saving, logging, target database control as well as managing exceptions and errors. 

The applications are defined as a list of actions for the automated operation of a MultiTelescope object. Since action modules are designed to initiate specific operations of SingleTelescope or MultiTelescope, users need to design a sequence of actions for autonomous observation by combining utility modules with action modules. With the flexibility of Python, such applications can be scheduled for a fully-robotic observation. For example, the default TCSpy application is shown in Section~\ref{subsec:7DT_implementation_Roboticobs}.

\subsection{Target database}
The Target database is a MySQL database that serves as a pivotal interface in TCSpy. It stores targets for nightly observations, dynamically updates their observing status, and calculates optimal targets throughout the observation. Before observation begins, an initialization process is triggered for a faster scoring process of optimal target selection. In this process, celestial information such as rise time, set time, and moon separation at the observing site is calculated with astroplan\cite{astroplan_2018} library. During observation, the observation scheduler selects the target in the best observing condition to trigger real-time observation. The selection process is based on a scoring algorithm. At first, the observability of the targets is checked. This process includes checking the moon separation and altitude of the targets. Second, scores of all targets are defined as a relative altitude and priority with the following equation.

\begin{equation}
\label{eq:fov}
Score = w_{alt}\times \frac{Altitude}{Altitude_{max}} + w_{prior}\times \frac{Priority}{Priority_{max}}
\end{equation}

The maximum altitude of a target is defined as the maximum altitude during the observing night. This simple algorithm offers flexibility and rapid calculation speed compared to more complex scoring algorithms. Users can easily prioritize desired targets by adjusting the weight(\begin{math}w\end{math}) and setting priorities according to specific needs. With this straightforward approach, 30,000 targets can be calculated in one second, as all necessary values are pre-calculated during the initialization process.

\section{Implementation of TCSpy on 7DT}
\label{sec:implementation_of_tcspy_on_7DT}
We implement TCSpy on 7DT to provide the various observation requirements provided in Section~\ref{subsubsec:observation_modes}. Running on MCC, TCSpy establishes robust connections with multiple telescope units, allowing commands to be sent simultaneously with the target database. Such a centralized control system enable the implementation of the following functionalities.

\subsection{Synchronized observation modes}
\label{subsec:7DT_implementatation_SyncObsmode}
As mentioned in Section~\ref{subsubsec:observation_modes}, concurrent operation of multiple telescope units is essential for spec and deep observation modes. These observation modes are executed within high-level action modules. Here, we’ll describe the spec observation mode, where different telescope units observe the designated target simultaneously using different medium-band filters. When the spec mode is triggered, TCSpy specifies each telescope to perform observations with specified filters from a pre-defined list of filters. This filter information is then passed to the level 0 Multiaction module to trigger the simultaneous operation of multiple telescope units. In the Multiaction module, TCSpy assigns a single core to each telescope unit with multiprocessing\cite{multiprocessing} library for simultaneous and expedited operation. For the default spec mode of 7DT, each telescope sequentially observes two medium-band filters to capture the entire spectrum. This approach enables observations on all 7DT units to be triggered within one second.

\subsection{Robotic observation}
\label{subsec:7DT_implementation_Roboticobs}
Robotic observation of multi-telescope array can be executed by running and scheduling applications. TCSpy supports built-in applications for the processes including startup, automatic flat acquisition, automatic night-time observation, and shutdown. Additionally, users have the flexibility to easily create custom applications for specific needs using action and utility modules. Here, we introduce the NightObservation application as an example.

The NightObservation application automatically triggers the observation of targets in the target database during the nighttime. Once triggered, it starts with the initialization process. The initialization process includes running the weather updater, initializing the target database, and checking device connections. Once the Sun falls below the horizon, the application schedules the most optimal target for observation from the target database every second. By comparing the required number of telescope units and available idle units in the telescope queue, the observation is triggered. The telescope queue and observing status of the target are promptly updated according to the progress of the observation, ensuring seamless operations. When the weather conditions become bad, all ongoing observations are aborted until the weather conditions improve. Based on the visibility of targets, aborted target acquisition resumes. This automated observation process continues until sunrise, providing fully automatic observation.

\begin{figure} [hb]
\begin{center}
\begin{tabular}{c} 
\includegraphics[height=11cm]{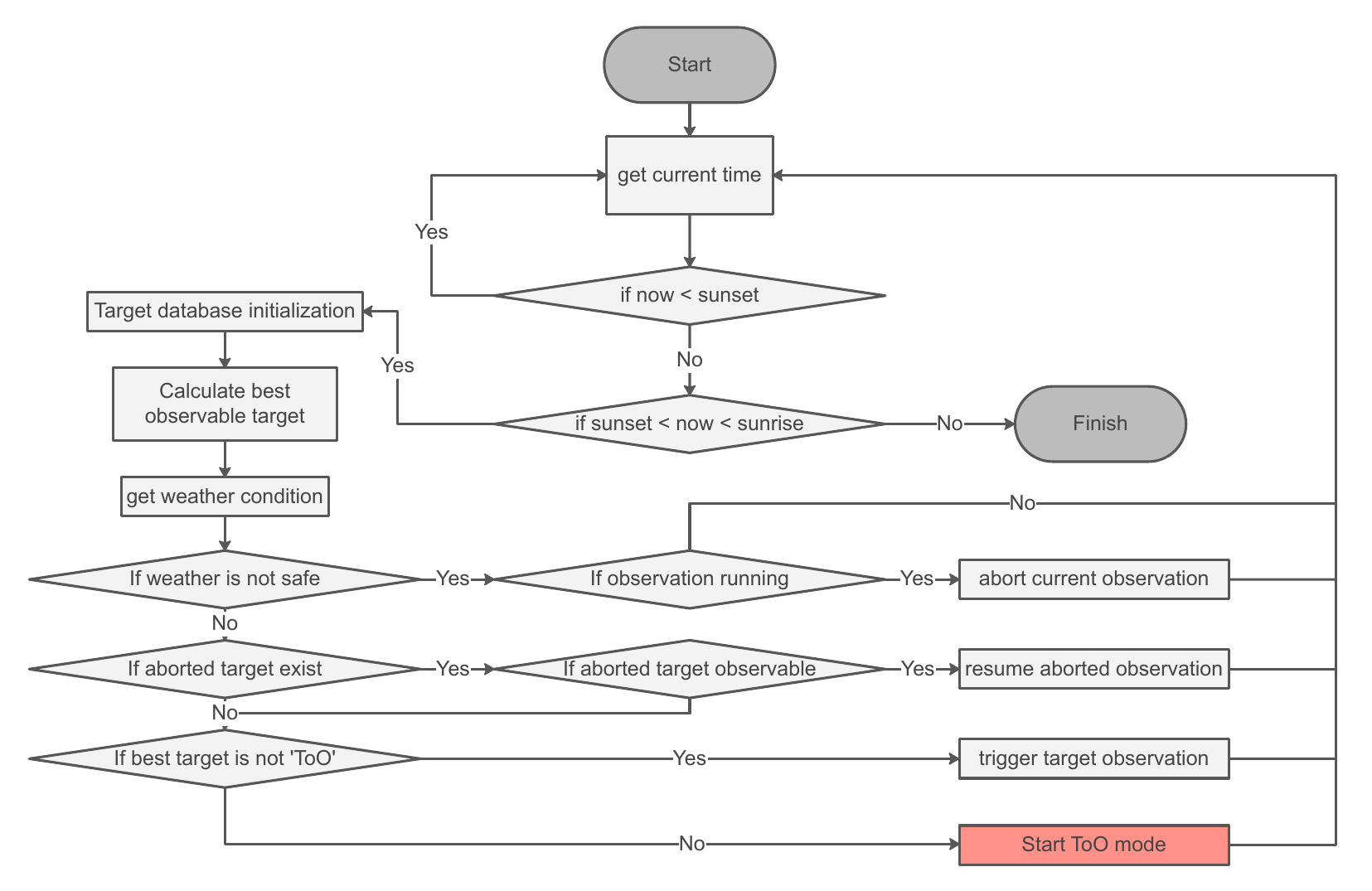}
\end{tabular}
\end{center}
\caption[example] 
{ \label{fig:video-example} 
The flowchart of the NightObservation application. With this application, 7DT triggers automatic observation of targets with the rapid responsiveness of ToO observation. For more detail, see Section~\ref{subsec:7DT_implementation_Roboticobs} and Section~\ref{subsec:7DT_implementation_rapidToO}.
}
\end{figure} 

\subsection{ToO Observation}
\label{subsec:7DT_implementation_rapidToO}
When ToO targets are received from authorized users or alert brokers, the ToO response can be carried out manually by users. Manual control by observers, however, increases the response time of ToO observation. Therefore, for a faster ToO response, the ToO implementation is incorporated into the NightOservation application discussed in Section~\ref{subsec:7DT_implementation_Roboticobs}. In the application, ToO targets have the highest priority when scheduling the most optimal target. When the ToO target is selected as the optimal target, the application interrupts all ongoing observations. This process can be rapidly prepared within 5 seconds, and all telescope units are ready to trigger ToO observation. Once the ToO observation concludes, observations of the previously halted targets resume.

\section{Conclusion}
This paper presents the fundamental structure, functionality, and application of TCSpy in controlling multi-telescope array. TCSpy establishes connections with multiple telescopes constructed with ASCOM-supported devices using the local network at the telescope site. Through this connection, TCSpy not only provides a multiple telescope control system but also offers various functionalities such as ToO observation and synchronized observation modes. 

In section~\ref{sec:implementation_of_tcspy_on_7DT}, we demonstrate the successful application of TCSpy to 7DT. With the high-level action modules, synchronized observation of multi-telescope array is implemented with multiprocessing. Moreover, the default application of TCSpy, NightObservation, effectively implements robotic observation with the ToO target and weather condition monitoring.  Finally, 7DT equipped with TCSpy can initiate simultaneous observations across all the telescope in the multiple telescope system in just one second, while preparing for ToO observations within five seconds.

\acknowledgments 
This work was supported by the National Research Foundation of Korea (NRF) grants, No. 2020R1A2C3011091, and No. 2021M3F7A1084525, funded by the Korea government (MSIT).

\bibliography{report} 
\bibliographystyle{spiebib} 

\end{document}